\documentclass[11pt,a4paper]{article}

\usepackage[T1]{fontenc}
\usepackage[utf8]{inputenc}

\usepackage{amsmath}
\usepackage{amssymb}

\usepackage{multirow}
\usepackage{makecell}
\usepackage{placeins}

\usepackage{lineno}

\usepackage{graphicx}
\usepackage{xcolor}

\usepackage{hyperref}

\usepackage[style=authoryear, maxnames=1]{biblatex}
\usepackage[margin=2.5cm]{geometry}

\newcommand{\ack}[1]{
\section*{Acknowledgements}
#1
}

\newcommand{\funding}[1]{
\section*{Funding}
#1
}

\newcommand{\roles}[1]{
\section*{Author contributions}
#1
}

\newcommand{\data}[1]{
\section*{Data and code availability}
#1
}

\newcommand{\orcid}[1]{\href{https://orcid.org/#1}{\includegraphics[width=8pt]{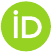}}}

\begin{document}
%\linenumbers
\title{Probabilistic storyline attribution using machine learning}

\author{
Frieder Loer\textsuperscript{1,*}\orcid{0009-0005-7440-0466} \and
Maybritt Schillinger\textsuperscript{2}\orcid{0000-0001-6763-3353} \and
Sebastian Sippel\textsuperscript{1}\orcid{0000-0002-4510-4458}
}

\maketitle

\begin{center}
\textsuperscript{1}Institute for Meteorology, Leipzig University, Leipzig, Germany\\
\vspace{0.2cm}
\textsuperscript{2}Seminar for Statistics, ETH Zurich, Switzerland\\
\vspace{0.2cm}
\textsuperscript{*}Corresponding author: \texttt{frieder.loer@uni-leipzig.de}
\end{center}

\vspace{0.5cm}


%TC:ignore
\begin{abstract}
A fundamental goal in climate attribution is to estimate how forced climate change contributes to observed extreme weather events. The storyline attribution method compares an observed weather event, conditional on its atmospheric dynamic state (i.e., atmospheric circulation), in the current, 'factual' climate to an event with very similar circulation conditions in a hypothetical, 'counterfactual' climate. However, physical climate models cannot directly transfer these storyline counterfactuals across different climate forcing states. 

Statistical and machine learning techniques may overcome this limitation; yet, emulating circulation-conditional extreme events under different climate states is challenging.
Here, we demonstrate distributional autoencoders (DAEs) as a versatile method for generating climate counterfactuals. They model the full distribution of spatially resolved European temperature fields conditional on the atmospheric circulation state and the mean global warming level. These distributions allow for deriving meaningful conditional probability ratios, which is a particular advantage of the DAE-based storyline approach.
We train DAEs on fully coupled climate model simulations and we evaluate the modelled distributions across different factual and storyline-based counterfactual climate model simulations. 
In an illustrative case study, we revisit the 2003 European heatwave and we generate counterfactuals for a hypothetical `2003-like European heatwave' using ERA5 circulation, which we hypothesize to occur a quarter century (2028) and a half century (2053) after 2003. The conditional intensity would increase from 29.3 °C in 2003, to 30.3 °C and 32.1 °C in 2028 and 2053, respectively and conditional probability ratios would be 2.1 and 3.2 when compared to 2003. Overall, this study introduces a new deep learning method that complements the toolbox for extreme event attribution. DAE-based storyline attribution may provide new avenues towards distributionally explicit generation of counterfactuals and could be extended in future studies towards attributing single forcing effects, or climate impacts.
\end{abstract}
%TC:endignore

\section{Introduction}
%%%%%%%%%%%%%%%%%%%%%
%%% General Intro %%%
%%%%%%%%%%%%%%%%%%%%%
The goal of extreme event attribution is to isolate the signal of anthropogenic forcings in weather events despite large internal variability \parencite{stott_human_2004, otto_reconciling_2012, peterson_explaining_2012}. Extreme event attribution thus aims to determine whether and how climate change has contributed to the frequency or intensity of specific events. Today, several methods attribute characteristics of individual weather events to forced climate change \parencite{thompson_need_2026}. Some methods are used quasi-operationally shortly after extreme events \parencite{philip_protocol_2020, faranda_climameter_2024} and the field appears to move toward attributing more complex events and impacts \parencite{zscheischler_attributing_2022, perkins-kirkpatrick_frontiers_2024}.

%%%%%%%%%%%%%%%%%%%%%%%%%%%%%%%%%%%%%%%%%
%%% Background risk-based attribution %%%
%%%%%%%%%%%%%%%%%%%%%%%%%%%%%%%%%%%%%%%%%
Extreme event attribution began with ideas and techniques now summarized under the terms probabilistic or risk-based attribution \parencite{Allen2003, stott_human_2004}. This approach combines extreme events with similar meteorological hazards into event classes and quantifies changes in their likelihood between the factual and counterfactual climate using large climate-model ensembles, often combined with extreme value analyses of observations \parencite{shepherd_common_2016, otto_attribution_2023, philip_protocol_2020}. A `counterfactual' describes the climate under a hypothetical alternative state, often the pre-industrial climate in event attribution, but potentially any other climate state \parencite{yiou_statistical_2017}. However, the risk-based approach has limitations in expressing the role of forced climate change in a \textit{particular event}: The event is defined via an event class \parencite{shepherd_common_2016}, which specifies the resulting hazard, but not the meteorological conditions leading to the specific event. This implies that probabilistic attribution may fall short when events are mainly driven by extreme dynamical conditions, which are often associated with high uncertainty under climate change \parencite{shepherd_common_2016}.

%%%%%%%%%%%%%%%%%%%%%%%%%%%%%
%%% Background Storylines %%%
%%%%%%%%%%%%%%%%%%%%%%%%%%%%%
These considerations have motivated storyline attribution: such approaches aim to isolate drivers of a particular extreme event, often by conditioning on its specific atmospheric circulation state while altering the background thermodynamic state of the atmosphere. Hence, event storylines analyse how a particular event would unfold in a counterfactual colder or warmer climate. Thereby, storylines isolate forced thermodynamic effects from those arising from changes in atmospheric dynamics. Because dynamical effects are often more variable, this increases the signal-to-noise ratio relative to unconditional attribution approaches \parencite{shepherd_common_2016}. Fixing the atmospheric circulation is motivated by the uncertain climate change signal in atmospheric circulation and the assumption that the corresponding circulation state could occur in a counterfactual climate due to internal variability. However, this disregards potential forced changes in atmospheric dynamics. Hence, probabilistic and storyline attribution approaches can be regarded as complementary and as spanning a spectrum of conditioning levels \parencite{thompson_need_2026, shepherd_common_2016}.

%%%%%%%%%%%%%%%%%%%%%%%%%%%%%%%%%
%%% Storyline Method Examples %%%
%%%%%%%%%%%%%%%%%%%%%%%%%%%%%%%%%
%Several techniques have evolved for modelling climate counterfactuals in event storyline settings, including approaches that condition on event dynamics. 
Several techniques have evolved for modelling climate counterfactuals that can be understood as event storylines, including approaches that condition on event dynamics.
\textit{Circulation analogues} statistically identify circulation states similar to the event of interest in model or reanalysis data and use them to estimate the hazard under similar circulation states in factual and counterfactual periods \parencite{yiou_statistical_2017, noyelle_attributing_2025}. However, closely controlling for the corresponding global mean temperature is difficult because it drastically limits sample size or analogue quality. Additionally, representing the analogue estimate by the mean of identified analogue hazard fields, as commonly done, tends to dampen extreme characteristics (regression to the mean). In other words, extreme events often lack good analogues because of their rarity.
\textit{Circulation-nudging} in climate models imposes the dynamical conditions of a transient climate simulation onto a simulation in a counterfactual background climate \parencite{feser_concept_2025, van_garderen_methodology_2021, sanchez-benitez_july_nodate, klimiuk_european_2025, pfleiderer2026}. The difference from the transient climate simulation can then be interpreted as an approximation of forced thermodynamic effects. However, it is difficult to provide a well-calibrated estimate of uncertainty across those simulations.

Furthermore, deep learning is increasingly studied in climate attribution because of its ability to learn non-linear climate signals and overcome limitations of methods relying on expensive numerical climate simulations. Deep learning has been used to directly model event storyline counterfactuals \parencite{trok_machine_2024, callahan_increasing_2026}. While demonstrating the potential of deep learning for estimating extreme event counterfactuals, these approaches do not explicitly model storylines spatially, evaluate predictions against counterfactual test data, or estimate conditional uncertainties.

%%%%%%%%%%%%%%%%%%%%%%%%%%%%%%%%%%%%%
%%% Gap: storylines + uncertainty %%%
%%%%%%%%%%%%%%%%%%%%%%%%%%%%%%%%%%%%%
Beyond the limitations of existing storyline methods, uncertainty in the meteorological hazard conditional on the event dynamics is rarely addressed \parencite{buschow_reconciling_2024, noyelle_attributing_2025}. A specific circulation state results in an expected temperature, but uncertainty remains due to variability in other temperature drivers, such as land-atmosphere interactions. Providing circulation-conditional uncertainty would strengthen attribution of changes in event intensity. More importantly, it would enable conditional probability ratios with well-calibrated uncertainty for specific meteorological hazards across climates given the event circulation. Few studies currently address this issue. For example, \cite{vignotto_towards_2020} statistically model the circulation-conditional temperature distribution. %however, the method does not strictly control for the background climate. 
\cite{noyelle_attributing_2025} derive conditional probability ratios in the analogue setting, and \cite{fonfay_combined_2026} provide storylines including conditional uncertainties at different warming levels, but each study with limitations as described above. 

%%%%%%%%%%%%%%%%%%%%
%%% Our Approach %%%
%%%%%%%%%%%%%%%%%%%%
Here, we present a generative deep-learning approach to event storylines that learns the relationship between atmospheric circulation, background warming, and temperature responses directly from fully coupled climate-model simulations. Our method jointly addresses three challenges not simultaneously resolved by current methods: 1) efficient transferability across climate-forcing states, 2) spatially resolved counterfactuals, and 3) well-calibrated uncertainty estimates via conditional temperature distributions.
Thus, the method addresses the question: 'What temperatures and uncertainty from non-circulation drivers can be expected for this circulation state in a counterfactual climate?' To achieve this, we leverage the 'engression' framework \parencite{shen_engression_2024}, a generative approach for modelling full conditional distributions that has previously been applied to rainfall-runoff modelling \parencite{kraft_modeling_2026} and multivariate downscaling \parencite{schillinger2026enscale}. We employ this framework in distributional autoencoders (DAEs, \cite{shen_distributional_2024}) to generate ensembles of spatially resolved temperatures with covariates on atmospheric dynamics and a proxy for background climate. Keeping the atmospheric circulation proxy fixed while varying the background-climate covariate enables generation of counterfactual conditional temperature distributions across climates. The resulting factual and counterfactual ensembles can then be analysed through conditional attribution.

Next, we introduce the data and methods used throughout the study. In the Results section, we comprehensively evaluate the method against factual and counterfactual CESM2 simulations (Subsection \ref{sec:model_eval}); and against a simple quantile-regression baseline (Subsection \ref{sec:baseline}). Finally, we showcase the framework for attributing extreme heat events for the 2003 European heatwave using ERA5 circulation covariates (Subsection \ref{sec:era5_hw}).

\section{Data and methods}

Our modelling approach shall generate ensembles of factual and counterfactual spatial temperature fields conditional on a given state of atmospheric circulation and a forced global mean temperature covariate. This covariate enables control of the background climate, i.e., to create event storyline counterfactuals across different climate forcing states. The generated ensembles represent conditional temperature distributions and thus quantify the temperature uncertainties resulting from variations in remaining drivers of temperature variability when conditioning on one circulation feature. 

\subsection{Data}

We establish our approach in a perfect model framework using the Community Earth System Model 2 Large Ensemble (CESM2-LE) \parencite{danabasoglu_community_2020, rodgers_ubiquity_2021}. 
% CESM2-LE
In addition, we use three individual, fully coupled and free running CESM2 simulations ($\mathrm{CESM2\text{-}ETH_{fact}}$) for testing, which were simulated under the same forcing boundary conditions as the large ensemble \parencite{pfleiderer2026, beyerle_2026_18172330}. 
% mention and explain nudging
These three simulations have corresponding circulation-nudged simulations ($\mathrm{CESM2\text{-}ETH_{cf\text{-}nudge}}$) \parencite{Bastos2023, singh_externally} that enable us to test the DAE-generated counterfactuals. The circulation-nudged simulations start from the identical initial conditions as the $\mathrm{CESM2\text{-}ETH_{fact}}$ simulations but with pre-industrial forcings and their horizontal wind fields are nudged towards the horizontal wind field of the free running, transient $\mathrm{CESM2\text{-}ETH_{fact}}$ simulations. Therefore, they can be regarded as samples from the circulation-conditional counterfactual temperature distribution that we aim to model with the DAE and provide well-suited test samples. %The free running and circulation-nudged simulations are described in more detail in \cite{pfleiderer2026}. 
% ERA5
Finally, we use ERA5 circulation data \parencite{hersbach_era5_2020}. We test the DAE-generated temperatures from ERA5 circulation against temperatures from two CESM2 simulations whose horizontal wind fields are nudged to the ERA5 horizontal wind field; one with historical and SSP370 forcings $\mathrm{CESM\text{-}ERA5_{fact\text{-}nudge}}$ and one with pre-industrial forcings $\mathrm{CESM\text{-}ERA5_{cf\text{-}nudge}}$. All circulation-nudged simulations are described in more detail in \cite{pfleiderer2026}. All datasets used throughout this study are given in Table \ref{tab1}.

We use 5-day mean values of Temperature ($\mathrm{T_{5d}}$) and detrended geopotential height at 500 hPa (Z500) to model European Temperatures from circulation in a North-Atlantic domain. For Z500, we compute its representation in the empirical orthogonal function (EOF) space using the first 1000 EOFs rather than working with the full field directly. The forced global mean temperature covariate (fGMT) is taken as the CESM2-LE mean global mean temperature (GMT) representing the response of global mean temperature to external climate forcing. The detailed data preprocessing is described in the SI.

We split the large ensemble (CESM2-LE) into 90 members for training and ten members for validation. The three factual ($\mathrm{CESM2\text{-}ETH_{fact}}$) and three counterfactual members ($\mathrm{CESM2\text{-}ETH_{cf\text{-}nudge}}$) are held out of the training process for the evaluation of the model as well as ERA5 data ($\mathrm{ERA5}$, $\mathrm{CESM2\text{-}ERA5_{fact\text{-}nudge}}$, $\mathrm{CESM2\text{-}ERA5_{cf\text{-}nudge}}$).

\begin{table}
\caption{Description of datasets used in this study. 
}
\centering
\footnotesize
\renewcommand{\arraystretch}{2}
\begin{tabular}{l l l l }
\hline
Dataset  & Description & Variables used & Purpose  \\
\hline
\hline
CESM2-LE  & \makecell[l]{100 climate simulations initialized with \\ perturbed initial conditions.} & Z500, T & \makecell[l]{Model training (90) /\\ validation (10)}  \\
\hline
\makecell[l]{$\mathrm{CESM2\text{-}ETH_{fact}}$} & \makecell[l]{3 free running CESM2 simulations \\ ($\mathrm{ETH_{fact}^{1300}}$, $\mathrm{ETH_{fact}^{1400}}$, $\mathrm{ETH_{fact}^{1500}}$)} & Z500, T & \makecell[l]{Test factual temperature}   \\
\hline
\makecell[l]{$\mathrm{CESM2\text{-}ETH_{cf-nudge}}$} & \makecell[l]{3 circulation-nudged CESM2 simulations \\ ($\mathrm{ETH_{cf}^{1300}}$, $\mathrm{ETH_{cf}^{1400}}$, $\mathrm{ETH_{cf}^{1500}}$)} & T & \makecell[l]{Test counterfactual \\ temperature}  \\
\hline
\makecell[l]{$\mathrm{CESM2\text{-}ERA5_{fact-nudge}}$} & \makecell[l]{CESM2 simulation nudged \\ to factual ERA5 horizontal winds.} & T & \makecell[l]{Heatwave case study \\ factual}   \\
\hline
\makecell[l]{$\mathrm{CESM2\text{-}ERA5_{cf-nudge}}$} & \makecell[l]{CESM2 pi-control simulation nudged \\ to factual ERA5 horizontal winds.} & T & \makecell[l]{Heatwave case study \\ counterfactual} \\
\hline
\makecell[l]{$\mathrm{ERA5}$} & \makecell[l]{ERA5 reanalysis product} & Z500 & \makecell[l]{Case study predictors}  \\
\hline
\end{tabular}
\label{tab1}
\end{table}

\subsection{Models and methods}

Our goal is to model the conditional distribution $p_{T \mid X}$, where $\mathrm{T_i} \in \mathbb{R}^{n \times n}$ (n: 32) denotes the full spatial temperature field for sample i and $X_i = (\mathrm{Z500_i}, \mathrm{fGMT_i})$ contains predictors describing detrended atmospheric circulation ($\mathrm{Z500_i} \in \mathbb{R}^{p}$) (p: 1000) and the climate forcing level ($\mathrm{fGMT}_i \in \mathbb{R}$).
In the following, we present a distributional autoencoder (DAE), a machine-learning approach that learns an approximation $\hat{p}_{T \mid X} \approx p_{T \mid X}$ %$q_{T \mid X} \approx p_{T \mid X}$, 
thereby enabling sampling from the conditional temperature distribution to generate an ensemble representing $\hat{p}_{T \mid X}$. In addition, quantile regression (QR) serves as a baseline.

\subsubsection{Generating temperatures with 'engression'}
\label{sec:engression}

Our approach is based on 'engression' \parencite{shen_engression_2024}, a generative framework developed for robustness in extrapolation tasks by modelling the full conditional distribution.
New data are generated by sampling from the modelled distribution using so-called 'pre-additive' noise that is concatenated with the predictors prior and during transformation by the employed model.

Following \cite{shen_distributional_2024, shen_engression_2024}, a suitable loss function is based on the energy score used in ensemble forecast evaluation \parencite{gneiting_2007} that involves a multivariate extension of the fair continuously ranked probability score \parencite{ferro_fair, leutbecher_fair}. 
In our conditional modelling task, a similar situation as in ensemble forecasting arises: We model a distribution $\hat{p}_{Y|X}$ and need to evaluate its quality against a single realization from the target distribution $Y_{i} \sim p_{Y|X_i}$. The energy score quantifies this using the sampled ensemble members $\hat{Y}_i^{(1)}, \hat{Y}_i^{(2)} \sim \hat{p}_{Y|X_i}$ and provides a suitable choice for our learning objective:

\begin{equation}
\mathcal{L}_{E}(Y,\ \{\hat{Y}^{(j)}\}_{j=1,2})
=
\frac{1}{N} \sum_{i=1}^N
\left[
\frac{1}{2} \sum_{j=1}^2
\left\lVert Y_i - \hat{Y}_{i}^{(j)} \right\rVert
-
\frac{1}{2}
\left\lVert \hat{Y}_{i}^{(1)} - \hat{Y}_{i}^{(2)} \right\rVert
\right].
\label{eq:loss_function_m2}
\end{equation}
\[
j: \text{ensemble member}
\]

This loss function is formulated for a two-member ensemble but can be extended for more members. The first term measures the reconstruction error between a generated sample and the target. The second term describes the expected distance ('variability') between two samples generated from the same predictor but different realizations of the pre-additive noise. $\left\lVert . \right\rVert$ is the Euclidean norm. This loss function describes a trade-off between accurate predictions and conditional uncertainty estimates.
Minimizing eq. \ref{eq:loss_function_m2} ensures that the generated ensemble follows the target distribution $p_{Y|X}$, yielding the narrowest calibrated distributions. Here, calibrated means that the modelled values sample variability correctly. % and do not systematically over- or underrsample typical or extreme values. 

\begin{figure}[t]
\centering
\includegraphics[width=\textwidth]{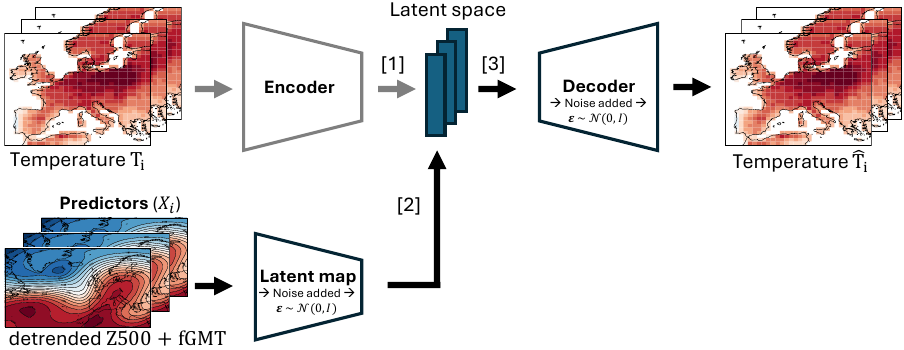}
\caption{Setup of the distributional autoencoder (DAE). The encoder reduces the dimensionality of the input temperature fields into the latent space and the decoder maps the latent elements back into the original feature space ([1]+[3]). The latent map regresses the latent elements onto the corresponding detrended $\mathrm{Z500}$ field and the $\mathrm{fGMT}$ covariate. Additionally, normally distributed ('pre-additive') noise is added inside the latent map and the decoder to make predictions generative (stochastic networks in navy). Only the latent map and decoder are used at inference time ([2]+[3]); the encoder [2] is only used for the training process (deterministic network in grey). Varying the $\mathrm{fGMT}$ covariate allows one to set the counterfactual thermodynamic background climate. With this setup, the model is expected to learn the conditional temperature distribution $Y \sim p_{T|X}$ with $X = (\mathrm{Z500}, \mathrm{fGMT})$.}
\label{fig:dae_model}
\end{figure}

\subsubsection{Distributional autoencoder setup}
\label{sec:dae_setup}

Using engression, we train a neural network with parameters $\theta$ that uses predictors $X$ and additional pre-additive Gaussian noise features $\varepsilon$, making the network stochastic and enabling generation of multiple ensemble members for each predictor set. The network defines a conditional distribution
\begin{equation}
    \mathrm{T} \sim \hat{p}_{T|X}, %\mid Z500, fGMT\right),
\end{equation}
where samples are generated via a stochastic mapping
\begin{equation}
    \mathrm{T} = f_{\theta}(\mathrm{Z500}, \mathrm{fGMT}, \varepsilon), \qquad \varepsilon \sim \mathcal{N}(0, I).
\end{equation}

Our setup is based on distributional autoencoders \parencite{shen_distributional_2024} (Figure \ref{fig:dae_model}). An encoder compresses high-dimensional temperature fields into a low-dimensional latent space, and a decoder reconstructs them into the original feature space. The encoder-decoder setup aims to reconstruct the input temperature fields as accurately as possible from their latent representations. Unlike traditional deterministic autoencoders, the decoder is stochastic, enabling sampling from the full conditional distribution of high-dimensional temperature fields given the latent-space value.
Following \cite{heinze-deml_latent_2021}, we additionally train a stochastic latent map that regresses latent-space elements $z$ onto the predictors $X = (\mathrm{Z500}, \mathrm{fGMT})$. The latent map captures uncertainty in latent temperature representations conditional on $\mathrm{Z500}$ and $\mathrm{fGMT}$.
Instead of learning a direct map from ($X, \varepsilon$) to T, we use the distributional autoencoder setup because separating predictor and target representations may improve flexibility for transfer learning. Moreover, this approach has proven more robust in multivariate modelling tasks \parencite{schillinger2026enscale}.

We jointly train all three components using the negative energy score (eq. \ref{eq:loss_function_m2}) as the learning objective. One training step consists of encoding and decoding a temperature field (path [1]+[3] in Figure \ref{fig:dae_model}), generating a latent element from $X$ using the latent map (path [2]), and decoding the predicted latent element (path [2]+[3]). For all three paths, we compute the empirical energy loss to update the model parameters. The encoder is only required during training to learn latent representations of the temperature fields. At inference time, only the path from predictors $X$ to the output is used ([2]+[3] in Figure \ref{fig:dae_model}).

The model is expected to generalize across temperature realizations of circulation states at different $\mathrm{fGMT}$ values, although not every circulation state occurs at every $\mathrm{fGMT}$ in the training data. Hence, we generate counterfactuals by varying the $\mathrm{fGMT}$ covariate while keeping circulation fixed. For fixed predictors, the stochastic latent map and decoder reproduce the conditional temperature variability. These variations represent temperature drivers beyond $\mathrm{Z500}$ and $\mathrm{fGMT}$.

\subsubsection{DAE architecture and training details}
\label{sec:dae_architecture}
Encoder, decoder and latent map consist of fully-connected multilayer neural networks. The latent map and decoder are 'stochastic networks' \parencite{shen_engression_2024} that concatenate a noise vector \textbf{$\varepsilon \sim \mathcal{N}(0, I)$} to each input and intermediate layer. We tune several model parameters; the procedure, selected configuration, and further training details are described in the SI. Importantly, for model selection, we evaluate the biases between DAE predictions and test set. Tuning results are broadly similar across a wide range of model configurations.

\subsubsection{Quantile regression as a statistical baseline}

We use linear quantile regression (QR) \parencite{koenker_regression_1978} as a distributional baseline for the DAE. QR estimates distributional quantiles rather than the conditional mean, as in ordinary least squares regression. The parameters of a quantile regression model are estimated by minimizing the quantile loss
\begin{equation}
\mathcal{L}(Y, \hat{Y}_{\tau}, \tau) =
\frac{1}{N} \sum_{i=1}^{N}
\begin{cases}
\tau \bigl(Y_i - \hat{Y}_{i, \tau}\bigr)      & \text{if } \hat{Y}_{i, \tau} \ge Y_i, \\
(\tau - 1)\bigl(Y_i - \hat{Y}_{i, \tau}\bigr) & \text{if } \hat{Y}_{i, \tau} < Y_i
\end{cases}
\label{eq:1}
\end{equation}
\noindent where $Y$ is the ground truth, $\hat{Y}_{\tau}$ the model-predicted quantile value, and $\tau \in (0,1)$ the nominal target quantile.

We fit the quantile regression model on the training dataset using identical predictors and one-dimensional targets given by aggregated temperatures over three regional domains (Spain, France, and Germany; see SI). We model quantiles $\tau \in {0.01, 0.02, ..., 0.99}$.

\subsection{Model evaluation}

We evaluate mean absolute error (MAE) and calibration of the generated temperature ensembles for the time series in each grid cell. We compute $\mathrm{MAE}=\frac{1}{N} \sum_{i=1}^{N} \left| Y_i - \hat{Y}_i \right|$ using for $\hat{Y}_i$ the median of the predicted DAE ensembles, allowing direct comparison with the conditional median predicted by the QR baseline. We quantify calibration using the mean absolute error (\(\mathrm{MAE}_{\mathrm{cal}}\)) between the model calibration curve and the perfect 1:1 line, following a similar notion of miscalibration as in \cite{wessel_enforcing_2026} and \cite{kraft_modeling_2026}. 
The calibration curve relates nominal quantile levels to empirical coverage, where the empirical coverage at level \(\tau\) is defined as $\frac{1}{N} \sum_{i=1}^{N} \mathbf{1}\left(Y_i \le \hat{Y}_{i,\tau}\right)$, and \(\hat{Y}_{i,\tau}\) denotes the predicted quantile at level \(\tau \in \{0.01, 0.02, \ldots, 0.99\}\) for sample \(i\). For a perfectly calibrated model, the empirical coverage equals the nominal quantile level for all \(\tau\), yielding a \(1{:}1\) calibration curve. For the DAE ensemble, the predicted conditional quantiles are estimated by computing the corresponding quantiles of the 100-member ensemble at each time step.

\subsection{Storyline attribution statements}
\label{sec:attribution_statements}
To attribute a specific event, factual and counterfactual ensembles are generated using the same circulation predictors but different $\mathrm{fGMT}$ values. For the factual ensemble, $\mathrm{fGMT}$ is set to the forced response at the event time step ($\mathrm{fGMT=fGMT_{fact}}$); for the counterfactual ensemble, it is set to zero ($\mathrm{fGMT=0}$), corresponding to no forced warming. The resulting temperature ensembles enable computation of the circulation-conditional intensity change and probability ratio. We calculate intensity change as the difference between the factual and counterfactual ensemble medians
\begin{equation}
\Delta I =
q_{0.5}\!\left(\hat{p}_{T \mid Z500,\; fGMT_{fact}}\right)
-
q_{0.5}\!\left(\hat{p}_{T \mid Z500,\; fGMT_{cf}}\right).
\end{equation}

\noindent
Here, \(q_{0.5}\) denotes the median and \(\hat{p}\) the empirical ensemble. The event probability conditional on circulation and forcing state is $\mathbb{P}(Y \geq Y_{th}|\mathrm{Z500, fGMT})$, where $Y_{th}$ denotes the extreme event threshold \parencite{shepherd_common_2016, yiou_statistical_2017}.
The probability ratio for exceeding a heatwave threshold $\mathrm{T_{HW}}$ is computed as
\begin{equation}
PR_{cond.} = \frac{\mathbb{P}(\mathrm{T \geq T_{HW}|Z500, fGMT_{fact}})}{\mathbb{P}(\mathrm{T \geq T_{HW}|Z500, fGMT_{cf}})}
\end{equation}

We approximate the conditional probabilities by the fraction of ensemble members exceeding the threshold; and use a Gaussian kernel density estimate to approximate the probability density function. We bootstrap this process 2000 times to estimate the sampling distributions of $\Delta I$ and $PR_{cond.}$.

\section{Results and discussion}
In this section, we evaluate the DAE on factual and counterfactual test data and compare its performance with the regional quantile-regression baseline. We then present an attribution case study by revisiting the 2003 European heatwave using ERA5 circulation data.

\subsection{Model evaluation}
\label{sec:model_eval}
Our underlying assumption is that the model can learn circulation-conditional forced climate change effects from transient climate simulations and generalize them across a range of $\mathrm{fGMT}$ values. Indeed, evaluation against factual test simulations and circulation-nudged simulations with pre-industrial boundary conditions shows that the model can remove thermodynamic forced effects conditional on the $\mathrm{Z500}$ state.

\begin{figure}
\includegraphics[width=\textwidth]{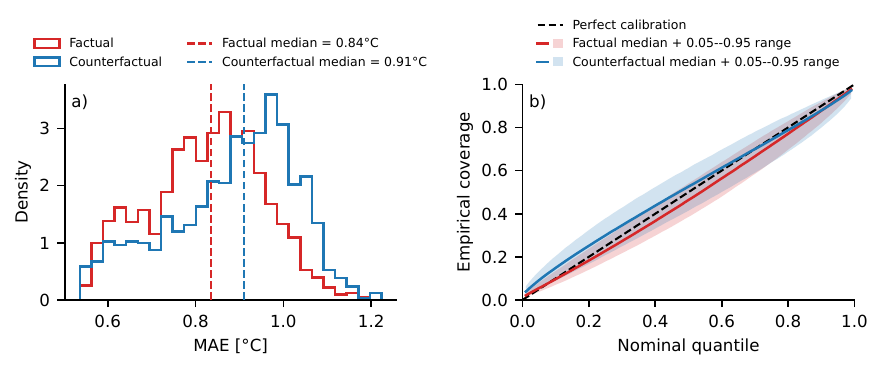}
\caption{\textbf{(a)} Distribution of MAE across all grid cells in the test set ($\mathrm{CESM2\text{-}ETH_{fact}}$, $\mathrm{CESM2\text{-}ETH_{cf-nudge}}$). \textbf{(b)} Conditional calibration of DAE ensembles evaluated against the corresponding test sets with shadings representing the spread and solid lines indicating the median across all grid cells.}
\label{fig:model_eval}
\end{figure}

The DAE yields comparable MAE distributions, with median values of 0.84°C and 0.91°C for the factual ($\mathrm{CESM2\text{-}ETH_{fact}}$) and counterfactual ($\mathrm{CESM2\text{-}ETH_{cf\text{-}nudge}}$) test sets, respectively (Fig.~\ref{fig:model_eval}a). 
The slightly higher counterfactual MAE may reflect that the $\mathrm{Z500}$ fields of the factual test members differ from those in the counterfactual circulation-nudged simulations. Calibration is close to the 1:1 line for both factual and counterfactual test sets, indicating that the DAE captures the conditional distributions well. This is a key goal in distributional regression and will be important for well-calibrated estimates of conditional probability ratios in the storyline attribution application (Subsection \ref{sec:era5_hw}). Although the factual and counterfactual calibration curves differ slightly (Figure \ref{fig:model_eval}b), their uncertainty ranges include the 1:1 line in both cases.

\subsection{Comparison to simple quantile regression baseline model on regional scale}
\label{sec:baseline}

\begin{figure}[h!]
\includegraphics[width=\textwidth]{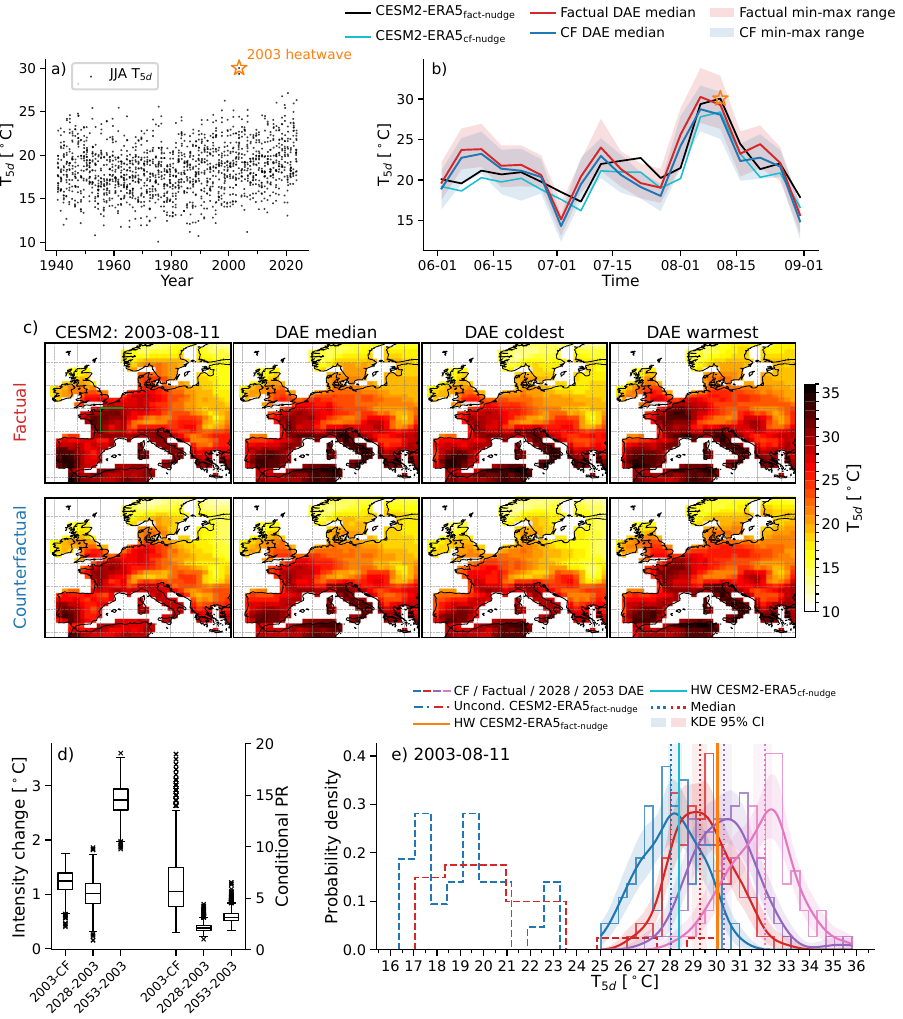}
\caption{\textbf{(a)} Absolute JJA $\mathrm{T_{5d}}$ temperatures averaged over the FR domain. The $\mathrm{T_{5d}}$ centred around 2003-08-11 is indicated by an orange star. \textbf{(b)} Time series of summer 2003 temperatures averaged over the green domain box shown in (c). \textbf{(c)} Factual and counterfactual events simulated by CESM2 (first column), with the remaining columns showing the corresponding DAE-generated ensemble. The green box defines the study domain. \textbf{(d)} Median intensity change and conditional probability ratios of the 2003 event compared to the counterfactual, 2028 and 2053 cases. Outliers of the 2003-CF conditional PR extend higher, but the axis is truncated for clarity. \textbf{(e)} Empirical factual and counterfactual distributions of DAE-generated, domain-averaged temperatures for factual and three counterfactual cases. Gaussian kernel density estimates represent the corresponding probability density functions. Shadings indicate the 95\% uncertainty interval obtained from bootstrapping. The unconditional temperatures on the 08-11 of each year from the test set ($\mathrm{CESM2\text{-}ERA5_{fact-nudge}}$) in a factual (1988-2018) and counterfactual (1940-1970) period are added as histograms.}
\label{fig:era5_hw}
\end{figure}

We compare our autoencoder against quantile regression (QR) model estimates of aggregated temperatures in a regional domain centred over France (Figure \ref{fig:era5_hw}c) to assess how DAE-derived quantiles compare with direct quantile estimates. Unlike the quantile regression models (see Methods), DAE temperatures are aggregated across grid cells only after ensemble generation. We focus on regional aggregates because our interest lies in regional climate characteristics rather than individual grid-cell temperatures. 

\begin{figure}[t]
\centering
\includegraphics[width=\textwidth]{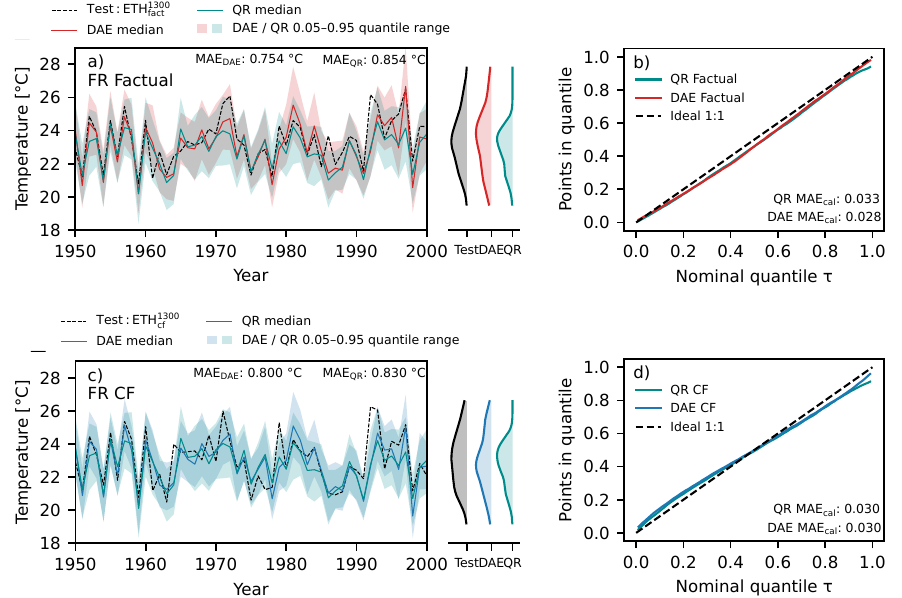}
\caption{Comparison of the DAE performance against a quantile regression model in the France domain (indicated in Figure \ref{fig:era5_hw}c). \textbf{(a)} Time series of the hottest 5-day period per year from: one member in the test set, the DAE model and the quantile regression model. Vertical densities show the distributions across the 1950-2000 period (QR density only represents the modelled median). \textbf{(b)} Calibration curve of the quantile regression model and the DAE model. The equivalent is shown for counterfactual temperatures in panel c) and d).}
\label{fig:qu_regr_comparison}
\end{figure}

Distributional regression involves several key strengths (Figure \ref{fig:qu_regr_comparison}): Extreme temperatures are often underestimated by the median predictions of both QR and DAE models but are captured by their predictive distributions. This reflects the well-known link between temperature and atmospheric circulation: extreme temperatures typically require exceptional circulation states. Consequently, circulation states associated with extremes tend to produce extreme temperatures. However, exceptionally hot temperatures are amplified by variability in drivers beyond $\mathrm{Z500}$. Our approach captures both extremes driven by unusual circulation and additional variability under a given circulation state.

Overall, visual inspection of Figure \ref{fig:qu_regr_comparison} reveals no systematic performance differences. However, over the analysed time period, the DAE yields lower MAE in both the factual and counterfactual cases. While this suggests better point predictions by the DAE, calibration differences are less clear (Figure \ref{fig:qu_regr_comparison}b,d). Results for two additional domains in the SI similarly show only minor differences from the MAE and calibration behaviour observed here. This is expected because QR directly models quantiles. Even under this conservative comparison, DAE-derived quantiles remain comparable to directly modelled quantiles, highlighting the model's ability to capture distributional features. This analysis highlights the aim of our approach: well-calibrated estimation of temperature uncertainty conditional on the $\mathrm{Z500}$ circulation feature. Overall, the DAE is competitive with the QR baseline while being more flexible and efficient: Instead of training separate QR models for each quantile and each region, only a single DAE must be trained. Moreover, the DAE generates full spatial fields, whereas QR only estimates quantiles of spatial aggregates.

\clearpage

\subsection{Revisiting the attribution of the 2003 European heatwave and potential future analogues}
\label{sec:era5_hw}

We revisit the 2003 European heatwave to illustrate a real-world application of DAE-based attribution.
For attribution, we use ERA5 $\mathrm{Z500}$ data from 2003-08-11 represented in EOF space and generate two 100-member DAE ensembles: a factual ensemble with the observed $\mathrm{fGMT}$ and a counterfactual ensemble with $\mathrm{fGMT}=0$, representing a pre-industrial climate (1850–1900). We compare the generated temperatures with CESM2 simulations whose horizontal winds are nudged to ERA5 under factual and counterfactual conditions ($\mathrm{CESM2\text{-}ERA5_{fact\text{-}nudge}}$ and $\mathrm{CESM2\text{-}ERA5_{cf\text{-}nudge}}$).
Applying a model trained on CESM2 circulation to the ERA5 circulation of 2003 may introduce a distributional shift. We therefore discuss its implications carefully below.

The 2003 heatwave produced exceptionally high temperatures in our study domain on 11 August 2003, as shown by $\mathrm{T_{5d}}$ in Figure \ref{fig:era5_hw}a. The early-August temperature peak is well captured by the ERA5-nudged CESM2 simulations (Figure \ref{fig:era5_hw}b). The influence of temperature drivers beyond $\mathrm{Z500}$ is reflected in deviations of factual and counterfactual temperatures from the DAE ensemble median (Figure \ref{fig:era5_hw}). These deviations are mostly captured by the DAE ensemble spread, highlighting its ability to represent temperature variability unexplained by $\mathrm{Z500}$. The DAE ensemble reproduces the spatial event characteristics well, although it slightly underestimates the eastward extent of the heatwave into Central Europe and overestimates temperatures in Italy and the coastal Maghreb (Figure \ref{fig:era5_hw}c).

Our DAE attribution suggests that these temperatures were about 1.3°C more intense and 5.7 times more likely in the climate of 2003 than in a pre-industrial climate under the same $\mathrm{Z500}$ state (Figure \ref{fig:era5_hw}d). %The factual intensity of the 2003 event is not reached in the counterfactual DAE ensemble, so we do not estimate a conditional probability ratio. However, transferring 
Transferring the event to hypothetical climates defined by the CESM2-LE forced response (fGMT) indicates that it would become 1°C and 2.7°C more intense, and 2.1 and 3.2 times more likely, in 2028 and 2053, respectively, relative to 2003 (Figure \ref{fig:era5_hw}e). Uncertainty arises from finite sampling of the 100-member ensembles and the bootstrap procedure. Uncertainties can be large for conditional probability ratios because the intense event is sometimes absent from the bootstrap sample or only contained a few times.

Consistent with our conditional analysis, the factual event lies in the tail of the histogram, while the $\mathrm{CESM2\text{-}ERA5_{cf\text{-}nudge}}$ simulation falls outside the range of observed temperatures (Figure \ref{fig:era5_hw}e).
This suggests that the event was largely dynamically driven but amplified by other sources of temperature variability. \cite{black_factors_2004} and \cite{garcia-herrera_review_2010} identify soil-moisture deficits and exceptional Mediterranean and North Atlantic sea-surface temperatures as contributors to the 2003 heatwave.
These results would be difficult to obtain with risk-based attribution alone. Our method isolates the climate change signal under the given dynamical conditions, which is a key advantage of storyline attribution \parencite{trenberth_attribution_2015, shepherd_common_2016}. Current storyline methods such as circulation nudging sample factual and counterfactual conditional distributions only sparsely, limiting similar inferences.

Although illustrated here for regionally averaged temperatures, the method provides spatially explicit attribution at grid-cell resolution, allowing arbitrary study domains and potential future applications in spatially explicit impact models. Nevertheless, the results rely on transferring the learned $\mathrm{Z500}$–temperature relationship from a single climate model. This may introduce a distribution shift between DAE-generated and ERA5 temperatures that reflects model–reanalysis differences. %However, these biases appear modest (Appendix Figure \hyperref[fig:era5_time_resolved_bias]{A11}) and could be further reduced through transfer learning in future work.
These biases (Figure \hyperref[fig:era5_time_resolved_bias]{S7}) could be further reduced through transfer learning in future work.
Nonetheless, our focus here is on differences between factual and counterfactual ensembles rather than their absolute temperatures \parencite{feser_concept_2025}.

\FloatBarrier

\subsection{Bridging storyline and probabilistic attribution with machine learning}

\subsubsection{The degree of conditioning and residual variability}
\label{sec:circulation_conditioning}

Existing \textit{storyline} attribution methods differ in the degree of conditioning on atmospheric circulation \parencite{feser_concept_2025, pfleiderer2026}. Deterministic statistical models, collectively termed \textit{dynamical adjustment} \parencite{feser_concept_2025}, typically impose strong circulation conditioning by relying on a specific set of predictors \parencite{smoliak_dynamical_2015, trok_machine_2024, cariou_linking_2025, saffioti_improved_2017, lehner_toward_2017, deser_forced_2016, sippel_uncovering_2019, terray_dynamical_2021}. In contrast, circulation analogue methods allow somewhat greater variability by comparing events with similar circulation states \parencite{noyelle_attributing_2025, yiou_statistical_2017}.

One interpretation is that conditioning on circulation implicitly constrains non-circulation drivers to states that co-occur with that circulation. \cite{pfleiderer2026} discuss this interpretation using examples of neural networks and circulation analogues in the context of land–atmosphere interactions. Consequently, the variance of circulation-conditional temperature distributions depends on the degree of conditioning. Broadly speaking, strong conditioning leaves little residual variability and thus no uncertainty estimate, whereas weaker conditioning permits greater variability in circulation and non-circulation drivers, resulting in larger but not necessarily well-calibrated uncertainty estimates.

Within this framework, the proposed DAE fills an important gap by directly modelling calibrated residual temperature variability under a given $\mathrm{Z500}$ state. The approach combines advantages of storyline methods with different conditioning strengths by enabling strong conditioning while providing calibrated conditional uncertainty estimates associated with remaining temperature drivers. At the same time, the DAE does not compromise point-prediction quality, as shown by comparison with the QR baseline (Section \ref{sec:baseline}). Uncertainty estimates arise from calibrated random noise in the DAE setup, and model evaluation shows that the generated variability matches that in the test data conditional on a given $\mathrm{Z500}$ state. Hence, the DAE captures hypothetical counterfactual variations in temperature drivers beyond $\mathrm{Z500}$, even when counterfactual states of other sources of temperature variability cannot be estimated explicitly.

\subsubsection{Relevance for extreme event attribution}
% AI, extreme events, climate attribution
Our study demonstrates the potential of deep learning for event storyline attribution, as also shown in previous studies \parencite{cariou_linking_2025, trok_machine_2024}.
However, we explicitly model the full circulation-conditional temperature distribution, thereby capturing the possibility of extreme temperatures under a given circulation state, rather than only the conditional mean as in most storyline methods (e.g. dynamical adjustment, circulation analogues, or nudged circulation).
To our knowledge, the DAE is the first generative deep-learning setup to explicitly model these uncertainties and evaluate them against counterfactual circulation-nudged storylines, and it also enables estimation of conditional probability ratios, complementing analogue-based approaches \parencite{noyelle_attributing_2025}.

% Transfer to real data
Ultimately, we envision this approach to provide highly efficient attribution of real-world heatwaves. Achieving this goal requires transferring the method from the perfect-model approach to reanalysis or observational data while addressing the limited numbers of observed extreme events. One strategy is to train on climate-model data and subsequently fine-tune on reanalysis data using transfer learning. The separation of target and circulation components may provide a flexible setup for such transfer tasks. Including multiple climate models during training, as in \cite{trok_machine_2024}, may further reduce model-specific biases. Further extensions may include higher spatial resolution and long-term temporal consistency of the residual noise. After successful transfer, the approach could move from analysing distributions of factual and counterfactual differences to the difference between factual and counterfactual distributions, a remaining limitation of storyline methods \parencite{feser_concept_2025}.

% potential for rapid attribution
Rapid attribution studies currently rely on risk-based \parencite{philip_protocol_2020} or circulation-analogue \parencite{faranda_climameter_2024} techniques. The proposed framework could serve as a complementary approach to these methods, as it conditions on the specific circulation feature of an event while enabling probabilistic statements through well-constrained conditional probability ratios. In principle, the approach is applicable to other climate variables such as precipitation and long-term events such as drought. The method could be especially valuable when circulation analogues are limited in number or quality, as is often the case for precipitation extremes. Given a suitable circulation domain and training dataset, the approach should be transferable to any region worldwide.

\section{Conclusion}
% summarize method
We have presented a framework for modelling counterfactual temperature distributions for event storyline attribution using distributional autoencoders (DAEs). During training, the DAE encodes and decodes temperatures while regressing latent space elements on atmospheric circulation and the climate background proxy, fGMT. The model includes stochastic features and is trained using the negative energy score. It thereby learns to generate temperature ensembles for a given circulation state across climate states. Evaluation against factual and counterfactual circulation-nudged simulations demonstrates high skill and well-calibrated conditional uncertainty estimates. %Comparison with a linear quantile regression model shows that the generated ensembles reproduce the expected distributional characteristics.

% summarize findings
This framework is developed in a perfect model setting as a proof of concept for extreme event attribution. We use the DAE to revisit the summer 2003 European heatwave in ERA5 over a domain centred on France using ERA5 $\mathrm{Z500}$ data. Given the 2003 circulation, our analysis suggests that anthropogenic forcing increased the intensity of the 2003 European heatwave by about 1.3°C and made it about 5.7 times more likely.
Translating the event to hypothetical climates representing 2028 and 2053 suggests that it would become 1°C and 2.7°C more intense, and 2.1 and 3.2 times more likely, relative to 2003. Although transfer from a perfect-model setting to reanalysis data implies a distributional shift and thus uncertainty, the results suggest that the model captures circulation–temperature relationships common to both climate models and reanalyses. Moreover, the method provides well-calibrated conditional probability ratios within a storyline framework, a capability not currently available in comparable attribution methods.

% methodological outlook
Given the limited record of deep learning in climate attribution, important questions remain regarding interpretability and the reliable representation of extreme events. Future research could therefore focus on understanding the representations learned by models, for example through explainable AI techniques \parencite{holzinger_xxai_2022, gomez-orellana_one_2023, yang_interpretable_2024, bommer_finding_2024}. In particular, this includes investigating whether models robustly capture distributional shifts between factual and counterfactual climates, and analysing the information encoded in the latent space \parencite{happe_detecting_2024, happe_interpretable_nodate, pacal_understanding_2025, carvalho-oliveira_targeted_2026}.

% general outlook
More generally, conditional distributional temperature modelling is not limited to event counterfactuals but can be applied wherever generative modelling of climate variables is required. For example, similar approaches could leverage high-resolution (nudged) climate simulations for regional climate emulation and projection. Such simulations are critical input for impact modelling but remain computationally expensive to generate with physical climate models. Another extension is to include additional covariates, such as aerosols or land-use changes, whose counterfactual values can be estimated. Single-forcing large ensembles, such as those from CESM2 \parencite{simpson_single_forcing}, provide a suitable framework for evaluating such experiments. This would broaden the range of attribution questions addressable with distributional modelling.

% final statement
Our study demonstrates a new avenue for extreme event attribution based on probabilistic deep learning. Future work should focus on transferring this approach to observational and reanalysis data to evaluate its potential under real-world conditions. We demonstrate how this can be done in a spatially explicit manner and across different climate states. We hypothesize that it could ultimately open a new avenue for rapid event attribution, not only due to its efficiency but also its ability to directly model circulation-conditional temperature distributions and resulting conditional probability ratios.

\clearpage

%TC:ignore

\ack{
We thank Urs Beyerle and István Dunkl for producing the nudged CESM2 simulations that were used to test our presented method. We thank all the scientists, software engineers and administrators who contributed to the development of the Community Earth System Model. We acknowledge the CESM2 Large Ensemble Community Project and supercomputing resources provided by the IBS Center for Climate Physics in South Korea. We acknowledge the use of ERA5 reanalysis data provided by the Copernicus Climate Change Service (C3S). Generative AI ChatGPT (GPT-5.5) was used for rephrasing sentences. 
}

\funding{
% This section is a list of funder names and grant numbers
Frieder Loer is funded by the Deutsche Forschungsgemeinschaft (DFG, German Research Foundation) via the graduate training school ECO-N (`Economics of Connected Natural Commons: Atmosphere and Biodiversity', GRK2939/1 – 506392361). Maybritt Schillinger is part of SPEED2ZERO, a Joint Initiative co-financed by the ETH Board. Sebastian Sippel acknowledges funding provided by the Heinz Maier-Leibnitz-Prize of the German Research Foundation and the EU Horizon project `Artificial Intelligence for Enhanced Representation of Processes and Extremes in Earth System Models' (AI4PEX; grant agreement 101137682).
}

\roles{
% List author names and the contributions made to the article, using terms from the NISO Contributor Roles Taxonomy (CRediT) https://credit.niso.org

Frieder Loer\orcid{0009-0005-7440-0466}

\noindent Conceptualization, Data Curation, Formal Analysis, Investigation, Methodology, Software, Visualization, Writing - original draft, Writing - review \& editing \\

\noindent Maybritt Schillinger\orcid{0000-0001-6763-3353}

\noindent Methodology, Software, Writing - review \& editing \\

\noindent Sebastian Sippel\orcid{0000-0002-4510-4458}

\noindent Conceptualization, Funding Acquisition, Investigation, Methodology, Project Administration, Resources, Supervision, Writing - review \& editing
}

\data{%The data used throughout this study is available at [insert link, not yet published]. 
The code used for producing the results is available at %\url{https://github.com/Motte12/Towards-a-distributional-autoencoder-for-climate-counterfactuals}.}
\url{https://github.com/Motte12/Towards-}
\allowbreak
\url{a-distributional-autoencoder-for-climate-counterfactuals}.}
% For more information on IOP Publishing's research data policy see: https://publishingsupport.iopscience.iop.org/questions/research-data/

%\suppdata{Sample text inserted for demonstration.}
%TC:endignore

\printbibliography
\clearpage

\section*{Supplementary Information}
\appendix
\setcounter{figure}{0}
\renewcommand{\thefigure}{S\arabic{figure}}

\section{Experimental setup}

\subsection{Study domain}

\begin{figure}[h]
\centering
\includegraphics[width=0.6\textwidth]{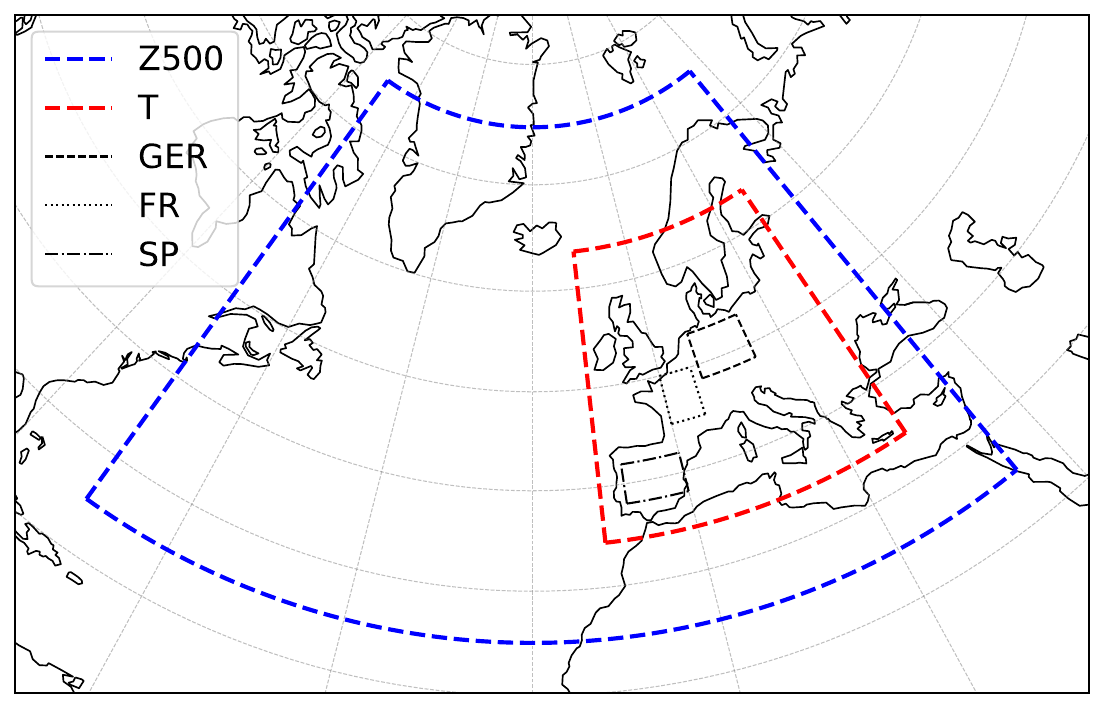}
\caption{Geopotential height at 500hPa (Z500, blue) and surface air temperature (T, red) domain. The Z500 domain spans the North Atlantic, Northern Africa, and Europe as this region represents atmospheric circulation influencing the temperature in the European domain. Subdomains for the quantile regression baseline comparison are indicated by black boxes.}
\label{fig:domain}
\end{figure}

\begin{table}[h!] % The optional [h!] parameter keeps the table close to where it's defined
\centering % Center the table horizontally
\begin{tabular}{|c|c|c|}
    \hline
     & Temperature & Z500 \\ \hline
    Extent & 34.4°N - 63.6°N, -11.25°W - 27.5°E & 25°N - 75°N, -70°W - 35°E \\ \hline
    Resolution & 1° $\times$ 1.25° & 1° $\times$ 1.25° \\ \hline
    Grid Size & $32 \times\! 32$  & $53 \times\! 85$ \\ \hline
\end{tabular}
\caption{Detailed description of the data domain. Latitude is $\approx$ 1°.} 
\label{tab:domain_extend} 
\end{table}

\subsection{Data preprocessing}

We model surface air temperatures (TREFHT, at 2m, only for $\mathrm{CESM2\text{-}ERA5_{fact-nudge}}$ it is TSA) in a European domain from geopotential height at 500 hPa (Z500) in a North-Atlantic domain (see Figure \ref{fig:domain}) and from a proxy for forced global mean temperature (fGMT). First, we compute 5-day non-overlapping mean values of temperature ($\mathrm{T_{5d}}$) and Z500 and select the time steps that fall into the summer months June, July, August (JJA). Per ensemble member, we compute seasonal anomalies w.r.t. the 1950-1980 climatology in the historical simulation (per 5-day time step). For temperature, we only use the land grid cells in the European domain by dropping grid cells that have less than 10\% land coverage. We detrend (the 5-day computed) Z500 on the grid cell level by subtracting the large ensemble mean to remove any potential first-order forced effects. We compute empirical orthogonal functions (EOFs) of the processed Z500 data (in JJA) and project each time step onto these EOFs to obtain a time series of our data in the form of EOF scores. We use the first 1000 scores to represent the Z500 data. We compute GMT as the weighted global average of the 5-day mean temperatures and compute the CESM2-LE mean to obtain the $\mathrm{fGMT}$ covariate. We standardize each predictor (including ERA5 Z500) by using the mean and standard deviation from the training set (90-members of CESM2-LE) computed over time. We project ERA5 Z500 onto the Z500 EOFs computed from CESM2-LE. The observationally derived ERA5 Z500 patterns are not detrended to preserve all data characteristics and because we are not evaluating any long-term trend characteristics in this paper. We train and evaluate our model on anomalies but display absolute temperatures in our results by adding the respective seasonal cycle back to the DAE model predictions.

\section{Distributional autoencoder architecture}

\subsection{Neural networks}
%\paragraph{Neural Network Architectures}
All components of the DAE were implemented as so-called 'stochastic networks' using the \texttt{StoNet} class of \cite{shen_engression_2024}, representing configurable multi-layer stochastic feedforward neural networks. Each network consists of \(n_L\) fully connected layers with hidden dimension \(n_d\), where \(n_L\) is required to be even. The layers are organized into residual blocks that each contain two affine transformations. Each affine transformation is followed by a ReLU activation function. Skip connections bypass every two layers by adding the transformed pathway to a residual shortcut connection. If the input and output dimensions differ, the shortcut is projected by a learned linear map; otherwise, the identity mapping is used. To introduce stochasticity, a Gaussian noise vector of dimension \(n_n\) is concatenated to the feature representation before each affine transformation. No activation function is applied at the output layer. The specific model components are as follows: 

\paragraph{Encoder} Maps input $T_i \in \mathbb{R}^{n \times n}$ (n: 32) to latent representations $\mathbf{z}_i \in \mathbb{R}^{n_{l}}$ ($n_l:$ latent dimensions) using $n_L$ layers, hidden dimension $n_d$, and a deterministic mapping (no noise added inside the network).

\paragraph{Decoder} Maps latent representations $\mathbf{z}_i \in \mathbb{R}^{n_l}$ back to reconstructed temperature fields $\hat{T}_i \in \mathbb{R}^{n \times n}$ using the same residual \texttt{StoNet} architecture as the encoder, with $n_L$ layers and hidden dimension $n_d$. In contrast to the encoder, the decoder permits stochasticity by concatenating Gaussian noise vectors of dimension $n_n$ before the affine transformations within each stochastic residual block.

\paragraph{Latent Map} Maps predictors ($X_i \in \mathbb{R}^{p}$) (p: 1001) to latent representations $\mathbf{z}_i \in \mathbb{R}^{n_{l}}$, using 2 layers, hidden dimension $n_{\mathrm{lm}}$, and noise dimension $n_{nlm}$.

\subsection{Model tuning}

We tune the model architecture and some training parameters across the values indicated in Table \ref{tab:hyperparameters} to identify the best configuration. The loss used in training (eq. \ref{eq:loss_function_m3}) is a linear combination of the energy loss (eq. \ref{eq:loss_function_m2}) evaluated for different components of the model (superscripts \textit{AE}, \textit{LM} and \textit{gen} indicate the model component to which the energy score is applied.). For each model configuration, we train the model (using Adam \parencite{adam_2017}) for 100 epochs and do not apply early stopping, as validation loss is not the only criterion for model selection. Instead, models are saved every ten epochs and the number of training epochs is included as a tuning parameter. Then, we perform a grid search to find the best model configuration. We evaluate each configuration on the ten-member validation set using the energy score, MAE and $\mathrm{MAE_{cal}}$. For tuning we only evaluate $\mathrm{MAE_{cal}}$ with quantiles $\tau \in \{0.05, 0.1, ..., 0.95\}$. In addition, we assess prediction bias on the test set to analyse how a bias in the factual setting transfers to the counterfactual setting. Finally, we rank all configurations across these metrics and select the model with the highest average rank. 
The training process itself is confined only to the factual, fully coupled simulations, hence there are no counterfactual simulations involved in the training process. During training, the empirical energy score is approximated using two generated samples per predictor, whereas 100-member ensembles are generated during inference.

\begin{equation}
L_{E}(Y,\ \{\hat{Y}^{(j)}\}_{j=1,2})
=
\frac{1}{N} \sum_{i=1}^N
\left[
\frac{1}{2} \sum_{j=1}^2
\left\lVert Y_i - \hat{Y}_{i}^{(j)} \right\rVert
-
\frac{1}{2}
\left\lVert \hat{Y}_{i}^{(1)} - \hat{Y}_{i}^{(2)} \right\rVert
\right].
\label{eq:loss_function_m2}
\end{equation}
\[
j: \text{ensemble member}
\]

\begin{equation}
\begin{aligned}
L(Y_i,\ X_i, \{\hat{Y}_{i}^{(j)}\}_{j=1,2})
&=
L^{AE}_{E}(Y_i,\ \underbrace{\{\hat{Y}_{i}^{(j)}\}_{j=1,2}}_{\hat{Y}^{(j)}_i:\ \mathrm{d(e}(Y_i))})
+
\lambda \, L^{LM}_{E}(\underbrace{z_i}_{z_i:\ e(Y_i)},\ \underbrace{\{\hat{z}_{i}^{(j)}\}_{j=1,2}}_{\hat{z}^{(j)}_i:\ \mathrm{lm}(X_i)})
+
\alpha \, L^{gen}_{E}(Y_i,\ \underbrace{\{\hat{Y}_{i}^{(j)}\}_{j=1,2}}_{\hat{Y}^{(j)}_i:\ \mathrm{d(lm}(X_i))})
% \\[0.3cm]
% &\qquad
% \hat{Y}^{(j)}_i:\ \mathrm{d(e}(Y_i))
% \qquad\qquad\qquad
% \hat{z}^{(j)}_i:\ \mathrm{lm}(X_i)
% \qquad\qquad\qquad
% \hat{Y}^{(j)}_i:\ \mathrm{d(lm}(X_i))
\\[0.3cm]
&\qquad
\mathrm{e()}:\ \mathrm{encoder \, transformation}
\\[0.3cm]
&\qquad
\mathrm{d()}:\ \mathrm{decoder \, transformation}
\\[0.3cm]
&\qquad
\mathrm{lm()}:\ \mathrm{latent \, map \, transformation}
\end{aligned}
\label{eq:loss_function_m3}
\end{equation}

\begin{table}[ht]
\centering
\caption{Hyperparameter tuning values with bold values indicating the selected configuration.}
\begin{tabular}{ll}
\hline
\textbf{Tuning Parameter} & \textbf{Values} \\
\hline
\multicolumn{2}{l}{\textit{Encoder and Decoder}} \\
\hline
Latent dimensions ($n_l$) & $\{50, \mathbf{100}\}$ \\
Hidden dimensions ($n_d$) & $\{50, \mathbf{100}\}$ \\
Number of layers ($n_L$) & $\{4, \mathbf{6}\}$ \\
Decoder noise dimensions ($n_n$) & $\{5, \mathbf{100}\}$ \\
\hline
\multicolumn{2}{l}{\textit{Latent map}} \\
\hline
Hidden dimensions ($n_{\mathrm{lm}}$) & $\{\mathbf{50}, 100\}$ \\
Noise dimensions ($n_{nlm}$) & $\{20, \mathbf{100}\}$ \\
\hline
\multicolumn{2}{l}{\textit{Training}} \\
\hline
$\lambda$ & $\{\mathbf{0.5}, 1.0\}$ \\
$\alpha$ & $\{\mathbf{1.0}, 1.5\}$ \\
Learning rate & $\{\mathbf{10^{-4}}, 5 \times 10^{-5}\}$ \\
Training epochs & $\{20, 30, ..., \mathbf{60}, ..., 100\}$\\
Batch size (not tuned) & 128 \\
\hline
\end{tabular}
\label{tab:hyperparameters}
\end{table}

\subsection{Model selection}
\label{sec:model_selection_results}

\begin{figure}[th!]
\centering
\includegraphics[width=\textwidth]{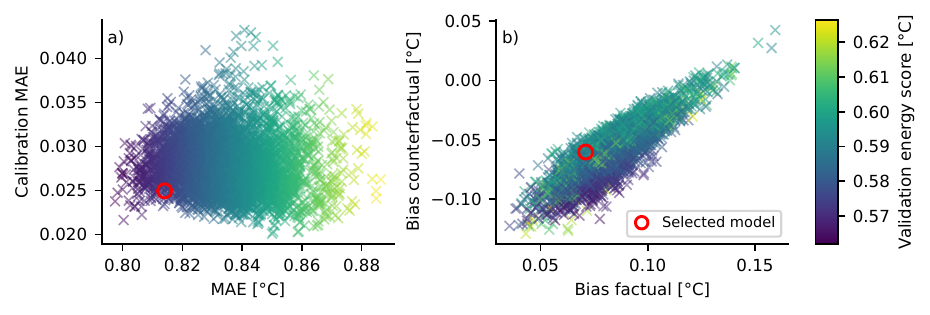}
\caption{Model selection results. The red circle indicates the selected model based on the best average rank among the selection metrics. The energy score on the validation set colorcodes the scatterpoints. \textbf{(a)} MAE versus calibration error ($\mathrm{MAE_{cal}}$) (averaged over all grid cell) on the validation set. \textbf{(b)} Factual versus counterfactual bias computed as the difference in mean of distributions of domain average temperatures truth - DAE) for the test set ($\mathrm{CESM2-ETH_{fact}}$ and $\mathrm{CESM2-ETH_{cf-nudge}}$ for factual and counterfactual respectively).}
\label{fig:model_selection}
\end{figure}

No clear relationship is apparent between $\mathrm{MAE}$ and calibration for all different fully trained models evaluated on the validation set as shown in Figure \ref{fig:model_selection}a. Yet, the energy score (colour labels in Figure \ref{fig:model_selection}a) seems to be dominated in first order by the $\mathrm{MAE}$ rather than by calibration. We further examine the relationship between the bias on the factual and counterfactual test sets as it is not immediately clear that a model trained only on factual training data will also perform well in the counterfactual case.
This is due to a potential distributional shift in the nudged circulation test data, that results from the nudging which is a physical model manipulation. A positive correlation for the biases between factual and counterfactual CESM2-ETH test sets emerges (Figure \ref{fig:model_selection}b), which implies that the difference between factual and counterfactual estimates across the fully trained DAE models appears approximately constant. This is a useful feature for the attribution applications presented in the main text. The selected configuration is indicated by a red circle in Figure \ref{fig:model_selection}. The selected model parameters are highlighted in bold in Table \ref{tab:hyperparameters}.

\section{Quantile regression model}

\subsection{Implementation of quantile regression models}

Here, we report the implementation of the quantile regression models used as distributional baselines. 
Due to the large size of the training dataset (429,210 data points with 1001 predictors and 1 predictand each), we implement quantile regression with a fully connected neural network consisting of one linear layer and no output activations. Hence, the network parametrizes each quantile as a linear combination of the predictors. Using stochastic gradient descent (Adam \parencite{adam_2017}), the network is trained to minimize the following loss: 
\begin{equation}
\mathcal{L}(Y, \hat{Y}, \tau) =
\frac{1}{N} \sum_{i=1}^{N}
\begin{cases}
\tau \bigl(Y_i - \hat{Y}_{i, \tau}\bigr)      & \text{if } \hat{Y}_{i, \tau} \ge Y_i, \\
(\tau - 1)\bigl(Y_i - \hat{Y}_{i, \tau}\bigr) & \text{if } \hat{Y}_{i, \tau} < Y_i.
\end{cases}
\label{eq:1}
\end{equation}
Here, $Y$ is the ground-truth, $\hat{Y}$ is the model prediction, and $\tau \in (0,1)$ specifies the target quantile. However, as this training strategy requires a differential loss but eq. \ref{eq:1} is not differentiable at $y - \hat{y} = 0$, we slightly smooth eq. \ref{eq:1} at $y - \hat{y} = 0$. The smoothing is described in the following. 

\begin{figure}[h]
\centering
\includegraphics[width=\textwidth]{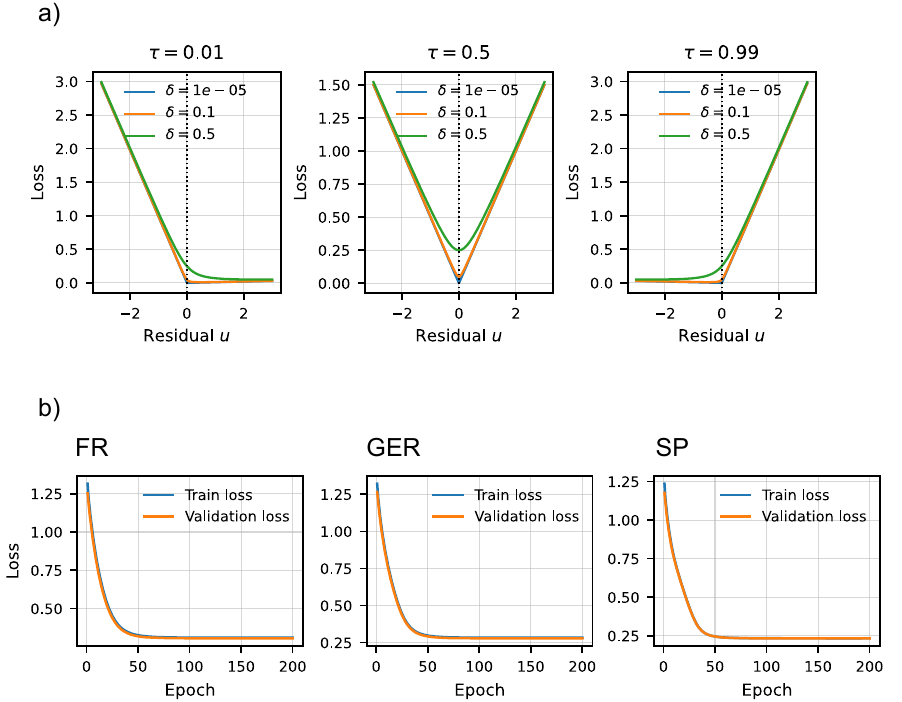}
\caption{Quantile regression model loss smoothing in a) and loss evolution during training in b).}
\label{fig:quantile_smoothed_loss}
\end{figure}

Starting from the residual
\[
u = y - \hat{y},
\]
the original quantile (pinball) loss is
\[
\rho_\tau(u)
=
\begin{cases}
\tau u, & u \ge 0, \\
(\tau - 1)u, & u < 0.
\end{cases}
\]
This can be rewritten as
\[
\rho_\tau(u)
=
\frac12
\left(
|u| + (2\tau - 1)u
\right).
\]
To obtain a differentiable approximation, the non-smooth absolute value term is replaced by
\[
|u|
\approx
\sqrt{u^2 + \delta^2},
\]
where \(\delta > 0\) controls the smoothing strength. This yields the smoothed quantile loss
\[
\rho_\tau^\delta(u)
=
\frac12
\left(
\sqrt{u^2 + \delta^2}
+
(2\tau - 1)u
\right).
\]
Substituting \(u = y - \hat{y}\) gives the final loss function used in this study. The effect of different smoothing parameter values $\delta$ is shown in Figure \ref{fig:quantile_smoothed_loss}a. Here, we select $\delta = 10^{-5}$ for training the QR model. We train the QR models for 100 epochs. The training and validation loss curves are shown in Figure \ref{fig:quantile_smoothed_loss}b. The fitted QR-model predicts all temperature quantiles per time step and corresponding predictors.

% \begin{table}
% \caption{Comparison of DAE and quantile regression (QR) model performance. Best values are highlighted in bold.}
% \centering
% \begin{tabular}{l ccc ccc ccc}
% \hline
%  & \multicolumn{3}{c}{\textbf{CRPS}} & \multicolumn{3}{c}{\textbf{R2}} & \multicolumn{3}{c}{\textbf{C-Q MAE}} \\
% \hline
%  & SP & FR & GER & SP & FR & GER & SP & FR & GER \\
% \hline
% \textbf{QR Factual} 
% & 0.231 & 0.305 & 0.280 
% & 0.917 & 0.872 & 0.892 
% & \textbf{0.015} & \textbf{0.033} & \textbf{0.038} \\
% \textbf{DAE Factual} 
% & \textbf{0.213} & \textbf{0.257} & \textbf{0.249} 
% & \textbf{0.930} & \textbf{0.913} & \textbf{0.917}
% & 0.020 & 0.037 & 0.055 \\
% \hline
% \textbf{QR CF} 
% & 0.249 & 0.303 & 0.294 
% & 0.793 & 0.797 & 0.807 
% & 0.023 & \textbf{0.030} & \textbf{0.038} \\
% \textbf{DAE CF} 
% & \textbf{0.241} & \textbf{0.292} & \textbf{0.291} 
% & \textbf{0.805} & \textbf{0.812} & \textbf{0.813}
% & \textbf{0.013} & 0.034 & 0.046 \\
% \hline
% \end{tabular}
% \label{table:baseline_comparison}
% \end{table}

\clearpage 
\section{Comparison of DAE to quantile regression models in SP and GER domains}

\begin{figure}[h]
\centering
\includegraphics{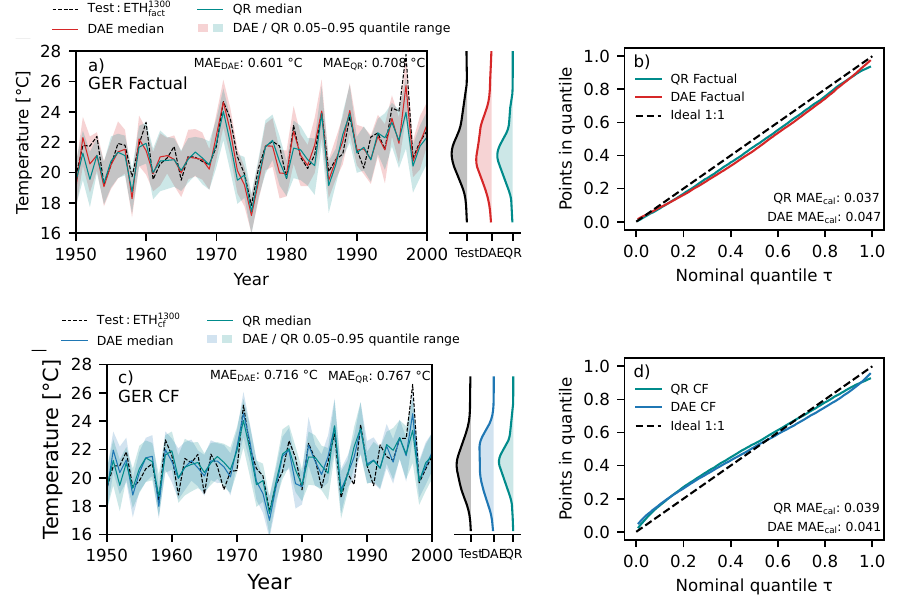}
\caption{Comparison of the DAE performance against a quantile regression model in the GER domain (indicated in Figure \ref{fig:domain}). \textbf{(a)} Time series of the hottest 5-day period per year from: one member in the test set, the DAE model and the quantile regression model. Vertical densities show the distributions across the 1950-2000 period (QR density only represents the modelled median). \textbf{(b)} Calibration curve of the quantile regression model and the DAE model. The equivalent is shown for counterfactual temperatures in panel c) and d).}
\label{fig:appendix_baseline}
\end{figure}

\begin{figure}[h]
\centering
\includegraphics{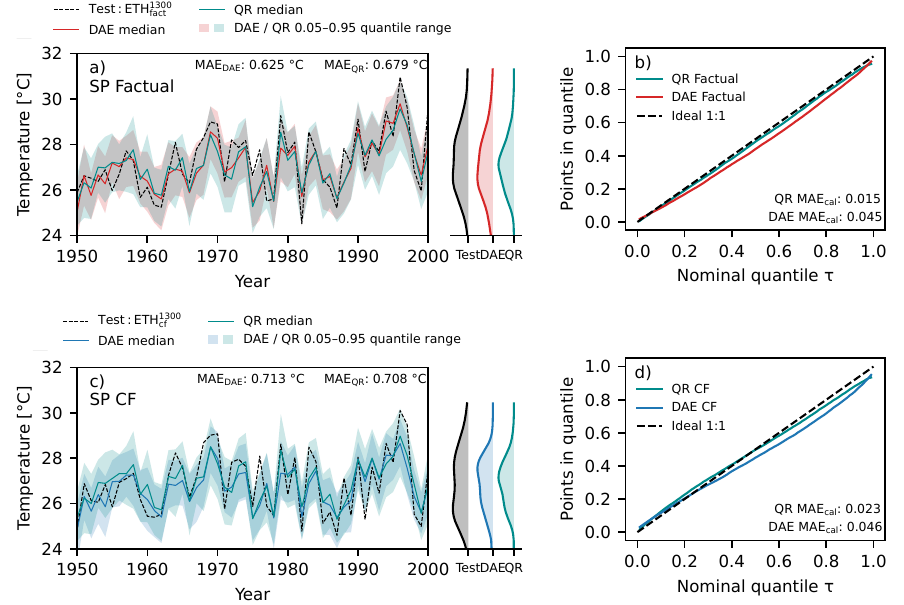}
\caption{Comparison of the DAE performance against a quantile regression model in the SP domain (indicated in Figure \ref{fig:domain}). \textbf{(a)} Time series of the hottest 5-day period per year from: one member in the test set, the DAE model and the quantile regression model. Vertical densities show the distributions across the 1950-2000 period (QR density only represents the modelled median). \textbf{(b)} Calibration curve of the quantile regression model and the DAE model. The equivalent is shown for counterfactual temperatures in panel c) and d).}
\label{fig:appendix_baseline}
\end{figure}
\clearpage

\section{An illustrative heatwave in CESM2}
\label{sec:synthetic_hw}

\begin{figure}[ht]
\includegraphics[width=\textwidth]{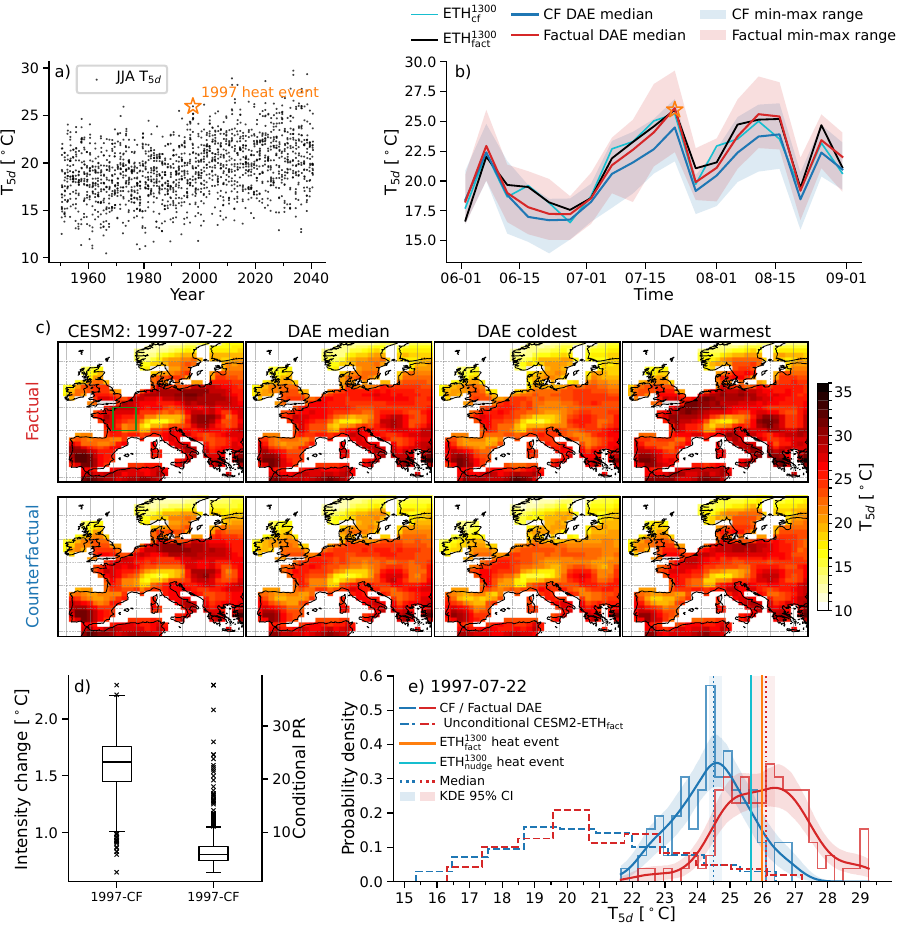}
\caption{\textbf{(a)} Absolute JJA $\mathrm{T_{5d}}$ temperatures averaged over the FR domain. The $T_{5d}$ centred around 1997-07-22 is indicated by an orange star. \textbf{(b)} Time series of summer 1997 temperatures averaged over the domain box shown in (c). \textbf{(c)} Factual and counterfactual events simulated by CESM2 (first column), with the remaining columns showing the corresponding DAE-generated ensemble. The green box defines the study domain. \textbf{(d)} Median intensity change and conditional probability ratio. \textbf{(e)} Empirical factual and counterfactual distributions of DAE-generated, domain-averaged temperatures, with Gaussian kernel density estimates of the corresponding probability density functions. Shadings indicate the 95\% uncertainty interval obtained from bootstrapping. The unconditional temperatures on the 07-22 of each year from the test set ($\mathrm{CESM2\text{-}ETH_{fact}}$) in a factual (1972-2022) and counterfactual (1850-1900) period are added as histograms.}
\label{fig:synthetic_hw}
\end{figure}

%TC:ignore
%We perform a storyline attribution case study of summer temperatures from the test data (test member $\mathrm{ETH_{fact}^{1300}}$) averaged in a regional domain centred over France around July 22, 1997. 
To illustrate an approximately unbiased storyline attribution, we include here a case study of summer temperatures from the test data (test member $\mathrm{ETH_{fact}^{1300}}$) averaged in a regional domain centred over France around July 22, 1997.
The intensity among summer temperatures is indicated in Figure \ref{fig:synthetic_hw}a. %\fl{
For attributing this event, we use the atmospheric circulation state of the corresponding date represented in the EOF space and create two 100-member DAE ensembles: one factual ensemble by setting fGMT to its actual value at the time and one counterfactual ensemble by setting $\mathrm{fGMT=0}$ representing a pre-industrial climate.
%}

Forced thermodynamic effects vary over the duration of the corresponding summer conditional on the corresponding circulation states which is represented by the difference of $\mathrm{ETH^{1300}_{fact}}$ and $\mathrm{ETH^{1300}_{cf}}$ summer 1997 temperature time series shown in Figure \ref{fig:synthetic_hw}b. Similarly, such varying offset is visible in the time evolution of factual and counterfactual DAE ensemble median illustrating how the DAE represents circulation conditional forced thermodynamic effects. The influence of temperature drivers besides $\mathrm{Z500}$ is again represented by the deviations of factual and counterfactual test temperatures from the modelled DAE ensemble median (Figure \ref{fig:synthetic_hw}b). However, these deviations are captured by the DAE ensemble spread, which highlights how residual temperature variability that remains unexplained from atmospheric circulation can be captured by the DAE distributional modelling. 
In addition to varying under different circulation states, forced thermodynamic effects also vary by region as is shown in Figure \ref{fig:synthetic_hw}c. The DAE-generated spatial temperatures show a clear separation of factual and counterfactual ensemble median, while the ranges from coldest to hottest member partly overlap in the factual and counterfactual ensembles from visual inspection (Figure \ref{fig:synthetic_hw}c).

Following our analysis, these temperatures were mainly circulation-driven. Yet, the counterfactual representation ($\mathrm{ETH_{cf}^{1300}}$, cyan vertical line in Figure \ref{fig:synthetic_hw}e) was further amplified by non-circulation temperature drivers, which is not the case for its factual representation that is located near the ensemble median (Figure \ref{fig:synthetic_hw}e). It is plausible that the relatively modest amplification of the factual representation may be explained by the fact that some soil moisture related feedbacks under extreme events \parencite{merrifield2019local} are not captured in the nudged circulation climate model simulations. Factual and counterfactual circulation-conditional temperature distributions give an intensity change of 1.6°C and a probability ratio of 5.8 for this temperature to occur under pre-industrial climate forcing given this particular circulation state as shown in Figure \ref{fig:synthetic_hw}d. The uncertainties in both quantities result from the limited sampling by the 100-member ensembles and the bootstrapping procedure.

Further, we compare the temperatures as simulated in the factual and counterfactual test set to the unconditional distributions of temperatures in the $\mathrm{CESM2\text{-}ETH_{fact}}$ on this day of the year in two respective 50-year periods (Figure \ref{fig:synthetic_hw}e). Conditioning on circulation increases the signal to noise ratio\footnote{\[
\mathrm{SNR}
=
\frac{|\mu_1 - \mu_2|}
{\sqrt{\frac{\sigma_1^2 + \sigma_2^2}{2}}}
\]

where $\mu_1,\mu_2$ are the means of the two distributions and
$\sigma_1,\sigma_2$ are their standard deviations.}
between factual and counterfactual of $0.14$ in the unconditional to $1.28$ in the conditional case, thereby isolating the forced thermodynamic effects due to  climate change \parencite{trenberth_attribution_2015, shepherd_common_2016}.
%TC:endignore

% \begin{table}[h!]
% \centering
% \caption{Intensity change and conditional probability ratio (PR) statistics for the CF scenario.}
% \begin{tabular}{lccc}
% \hline
% Metric & Median & Mean & 95\% CI \\
% \hline
% Intensity Change
% & 1.621
% & 1.593
% & [1.085, 1.969] \\

% Conditional PR
% & 5.824
% & 6.396
% & [3.363, 12.556] \\
% \hline
% \end{tabular}
% \label{tab:cf_stats}
% \end{table}
\FloatBarrier

\section{ERA5 bias in FR domain}

\begin{figure}[h!]
\centering 
\includegraphics[width=0.6\textwidth]{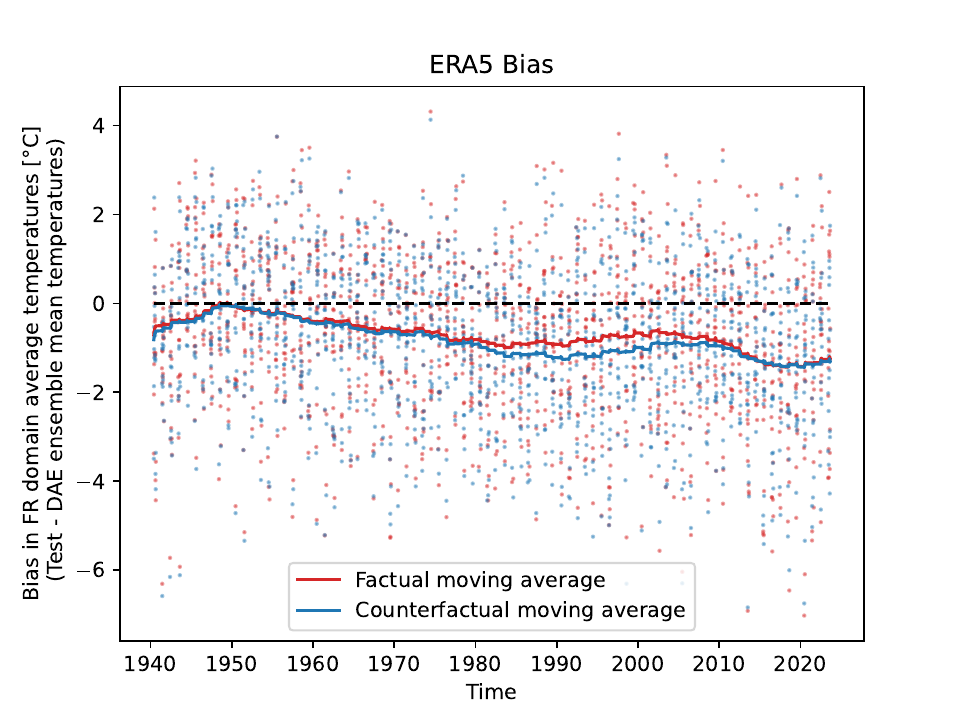}
%\caption{ERA5 time resolved bias. The bias is calculated as the difference of weighted domain average temperature between the true test set and the DAE ensemble mean for the factual and counterfactual case respectively. Scatterpoints show the exact difference and the solid lines show 200-point centered moving average with edge correction.}
\caption{ERA5 time resolved bias. The bias is calculated as the difference of weighted FR domain (see Figure \ref{fig:domain}) average temperatures between the true test set ($\mathrm{CESM2\text{-}ERA5_{fact\text{-}nudge}}$, $\mathrm{CESM2\text{-}ERA5_{cf\text{-}nudge}}$) and the DAE ensemble mean for the factual and counterfactual case respectively. Scatter points show the exact difference and the solid lines show 200-point centred moving average with edge correction.}
\label{fig:era5_time_resolved_bias}
\end{figure}

\FloatBarrier

\end{document}